\documentclass[conference]{IEEEtran}
\IEEEoverridecommandlockouts
\usepackage{cite}
\usepackage{amsmath,amssymb,amsfonts}
\usepackage{algorithmic}
\usepackage{graphicx}
\usepackage{textcomp}
\usepackage{xcolor}
\usepackage{multirow}
\usepackage{booktabs}
\usepackage{pifont}
\usepackage{hyperref}
\newcommand{\cmark}{\ding{51}}
\newcommand{\xmark}{\ding{55}}
\def\BibTeX{{\rm B\kern-.05em{\sc i\kern-.025em b}\kern-.08em
    T\kern-.1667em\lower.7ex\hbox{E}\kern-.125emX}}
\begin{document}

\title{EP-GAT: Energy-based Parallel Graph Attention Neural Network for Stock Trend Classification\\
%\thanks{This work was supported by the Royal Academy of Engineering.}
}

% \author{\IEEEauthorblockN{Anonymous Authors}}
\author{\IEEEauthorblockN{1\textsuperscript{st} Zhuodong Jiang}
\IEEEauthorblockA{\textit{School of Computer Science} \\
\textit{University of Bristol}\\
Bristol, UK \\
ci21041@bristol.ac.uk}
\and
\IEEEauthorblockN{2\textsuperscript{nd} Pengju Zhang}
\IEEEauthorblockA{\textit{School of Computer Science} \\
\textit{Univeristy of Bristol}\\
Bristol, UK \\
qo22685@bristol.ac.uk}
\and
\IEEEauthorblockN{3\textsuperscript{rd} Peter Martin}
\IEEEauthorblockA{\textit{School of Physics} \\
\textit{Univeristy of Bristol}\\
Bristol, UK \\
peter.martin@bristol.ac.uk}
}

\maketitle

\begin{abstract}
Graph neural networks have shown remarkable performance in forecasting stock movements, which arises from learning complex inter-dependencies between stocks and intra-dynamics of stocks. Existing approaches based on graph neural networks typically rely on static or manually defined factors to model changing inter-dependencies between stocks. Furthermore, these works often struggle to preserve hierarchical features within stocks. To bridge these gaps, this work presents the Energy-based Parallel Graph Attention Neural Network, a novel approach for predicting future movements for multiple stocks. First, it generates a dynamic stock graph with the energy difference between stocks and Boltzmann distribution, capturing evolving inter-dependencies between stocks. Then, a parallel graph attention mechanism is proposed to preserve the hierarchical intra-stock dynamics. Extensive experiments on five real-world datasets are conducted to validate the proposed approach, spanning from the US stock markets (NASDAQ, NYSE, SP) and UK stock markets (FTSE, LSE). The experimental results demonstrate that EP-GAT consistently outperforms competitive five baselines on test periods across various metrics. The ablation studies and hyperparameter sensitivity analysis further validate the effectiveness of each module in the proposed method. The raw dataset and code are available at \url{https://github.com/theflash987/EP-GAT}.
\end{abstract}

\begin{IEEEkeywords}
stock trend forecasting, graph neural network, dynamic graph modelling, graph attention, Boltzmann distribution.
\end{IEEEkeywords}

\section{Introduction}
\indent
Stock movement prediction remains a challenging problem for industry and academia, due to the stochastic nature~\cite{kasa1992common} and high volatility of the markets~\cite{adam2016stock}. It has been the central focus, as an insightful forecast about future movements provides valuable information for investors and policymakers~\cite{jiang2021applications}. Although extensive efforts have been made in this field, they often fail to model stocks' long-term temporal patterns~\cite{ganesan2021stock}. For instance, traditional forecasting models, such as LASSO linear regression~\cite{roy2015stock} and ARIMA~\cite{ariyo2014stock}, assume independence between stocks, neglecting intricate interdependencies between stocks and failing to generalize to longer time periods~\cite{kamble2024design}. Furthermore, according to previous studies~\cite{sapankevych2009time}, these models are solely valid for a single stock over a certain period. Recently, deep learning has become a preferred approach to open problems in various fields, such as financial markets~\cite{shi2022state,shi2023neural}, medical imaging~\cite{tian2022optimal,tian2024tagat}, navigation~\cite{yang2023conditional}, and social inference~\cite{bo2020social}. At the early stage, recurrent neural networks (RNNs), especially methods built on long short-term memory (LSTM), such as PCA-LSTM~\cite{wen2020research} and Bi-Directional LSTM~\cite{sunny2020deep} are applied to capture nonlinear temporal features in extended time series for stock price forecast~\cite{moghar2020stock}. Despite their advancements, these models overlook the inter-relations between stocks, which is a crucial factor affecting the dynamics of stocks~\cite{adheli2014intra}. Furthermore, predicting the exact stock prices is undesirable and impractical in real-world scenarios~\cite{gandhmal2019systematic}, where fast adaptation to market changes is required. Therefore, research focuses have shifted to forecasting future movements of stocks. 

\indent
Among various deep learning models, Transformer-based approaches achieve excellent performance predicting stock movements~\cite{zhang2022transformer}. The Transformer proposes an attention mechanism that can adaptively assign weights between elements in sequences~\cite{vaswani2017attention}. Compared to RNN-based approaches, these methods implicitly model the inter-relations between multiple stocks with their historical indicator series. For example, FDG-Transformer~\cite{li2022stock} integrates the frequency decomposition into the Transformer to capture the distinctive information of stocks belonging to the same cluster. Similarly, TEANet~\cite{zhang2022transformer}, an encoder-based framework captures the temporal dependencies by incorporating textual information like posts and news. Nevertheless, these methods fail to explicitly model inter-dependencies between stocks that might omit certain critical inter-relations between stocks~\cite{li2022stock}. The graph neural networks (GNNs) learn meaningful representation from the graphs by explicitly considering the inter-dependencies between nodes and node features via adjacency matrices and node feature matrices~\cite{patel2024systematic}. Therefore, GNNs become the promising solutions to stock movement forecast, which models multiple stocks as a graph~\cite{zhang2022research,you2024multi}. For instance, HyperStockGAT~\cite{sawhney2021exploring} utilizes pre-defined industry-related information (i.e. companies within the same sector are treated as connected) to construct stock graphs and learn the temporal patterns with the temporal attention mechanism. Moreover, GCN-LSTM~\cite{shi2024integrated} combines knowledge graphs and LSTM to model the complex relations between stocks. However, these approaches typically rely on static and invariant factors to generate stock graphs, limiting their ability to capture dynamic changes in the markets~\cite{song2023stock}. As former studies~\cite{li2021modeling, xiang2022temporal,you2024multi,qian2024mdgnn} show generating stock graphs with historical stock indicators time series is more effective than utilizing static industry relations, facilitating the modelling of the stochastically changing relations between stocks. Furthermore, previous studies~\cite{mantegna1999hierarchical,sawhney2021exploring,you2024dgdnn,you2024multi} demonstrate that preserving hierarchical intra-stock features prevents distortion of learned temporal features in the learned latent representations.

\indent
With these in mind, this work presents a novel framework, the Energy-based Parallel Graph Attention Neural Network (EP-GAT) for stock trend classification. It comprises a dynamic stock graph generation and a parallel graph attention mechanism. First, the stock graph is generated based on the Boltzmann distribution and energy differences between historical indicator series of stocks, which enables the modelling of the evolving inter-stock dynamics. Subsequently, the distinctive hierarchical intra-stock features are captured by the parallel graph attention mechanism, allowing the model to preserve latent representations from different layers. Consequently, EP-GAT can consider inter-stock relations and intra-stock features simultaneously, facilitating the representation learning of complex interactions of stocks. This work contributes in threefold:
\begin{itemize}
    \item We propose an original method for generating dynamic stock graphs based on the Boltzmann distribution and energy differences. The pair-wise connectivity is obtained by formulating the energy difference between stocks and the lag window as variables of the Boltzmann distribution.
    \item We present a novel parallel graph attention mechanism, which learns the hierarchical intra-stock features by preserving latent representations from different layers.
    \item We conduct extensive experiments on real-world datasets with 503 stocks across five years. Experimental results empirically show that EP-GAT consistently outperforms five competitive baselines over five datasets regarding three evaluation metrics. The ablation studies and hyper-parameter experiments further validate the effectiveness of the proposed framework.
\end{itemize}

\section{Related Work}

\subsection{Multivariate Time Series Modelling}
\indent
Multivariate time series modelling requires consideration of the interdependencies between multiple time series and their temporal features. For example, MCTAN~\cite{ren2022mctan} applies a channel attention mechanism to weigh different contributions of channels, with multi-head local attention to extracting the long-term temporal relations. Similarly, MR-Transformer~\cite{zhu2023mr} adopts a long short-term transformer to capture temporal dependencies, and a temporal convolution module to extract latent features. Moreover, MagicNet~\cite{luo2025magicnet}, proposes a method to consider both temporal and spatial dependencies with financial documents. It includes a text memory slot for each stock to aggregate the interactive influences among stocks. Additionally, CATN~\cite{he2022catn} leverages the tree structure to capture the inter-series correlations. Then, it learns temporal intra-series patterns through the multi-level dependency learning mechanism. Despite their strengths in extracting the interrelations, they fail to explicitly model such intricate and dynamic patterns. 

\subsection{Graph Neural Networks in Stock Movement Prediction}
\indent
GNNs have been widely applied in stock movement predictions, arising from their powerful ability to learn inter-relations between stocks and intra-dynamics of stocks. For example, DGDNN~\cite{you2024dgdnn} and MGDPR~\cite{you2024multi} utilize mutual information and information entropy with historical stock indicator time series to generate dynamic stock graphs, modelling both inter-stock dependencies and intra-stock features. HyperStockGAT~\cite{sawhney2021exploring} adopts static industry factors to construct the stock graph and proposes a hyperbolic graph attention on Riemannian manifolds. Additionally, AD-GAT~\cite{cheng2021modeling} generates the stock graph based on the supply-chain relationships. It employs element-wise multiplication to the attributes of the source firms and an unmasked attention mechanism to infer the dynamic firm relations. Yet, they often assume the inter-stock relations are static and neglect the hierarchical features within stocks.

\section{Preliminary}
\subsection{Notations}
\indent
In this work, nodes represent stocks while edges represent node-wise inter-dependencies, governed by a temporal graph $\mathcal{G}_t$ at time step $t$. It is characterized by an adjacency matrix $\mathbf{A}_t \in \mathbb{R}^{N \times N}$ and a node feature matrix $\mathbf{X}_t \in \mathbb{R}^{N \times (\tau F)}$, where $N$ denotes the number of stocks, $\tau$ denotes the lag window size, and $F$ denotes the number of stock indicators. Since this work performs the future trends classification of multiple stocks, the mapping relation can be formulated as the following form,
\begin{equation}
    f: \Omega_L(\mathbf{X}_t, \mathbf{A}_t) \rightarrow \mathbf{C}_t.
    \label{mapping}
\end{equation}
Here, $\Omega_L(\cdot)$ denotes the model EP-GAT with $L$ propagation layers, $\mathbf{C}_t \in \mathbb{Z}^{N \times \phi  \alpha}$ denotes the label matrix, $\phi$ denotes the forecast step, and $\alpha$ denotes the number of trends. Concretely, the model utilizes historical stock indicators from interval $[t-\tau + 1, t]$ to predict future trends of multiple stocks over interval $[t+1, t+\phi]$.

\begin{figure*}[tbhp]
\centering
  \includegraphics[scale=0.7, trim=5.5cm 4.5cm 6cm 4.5cm, clip]{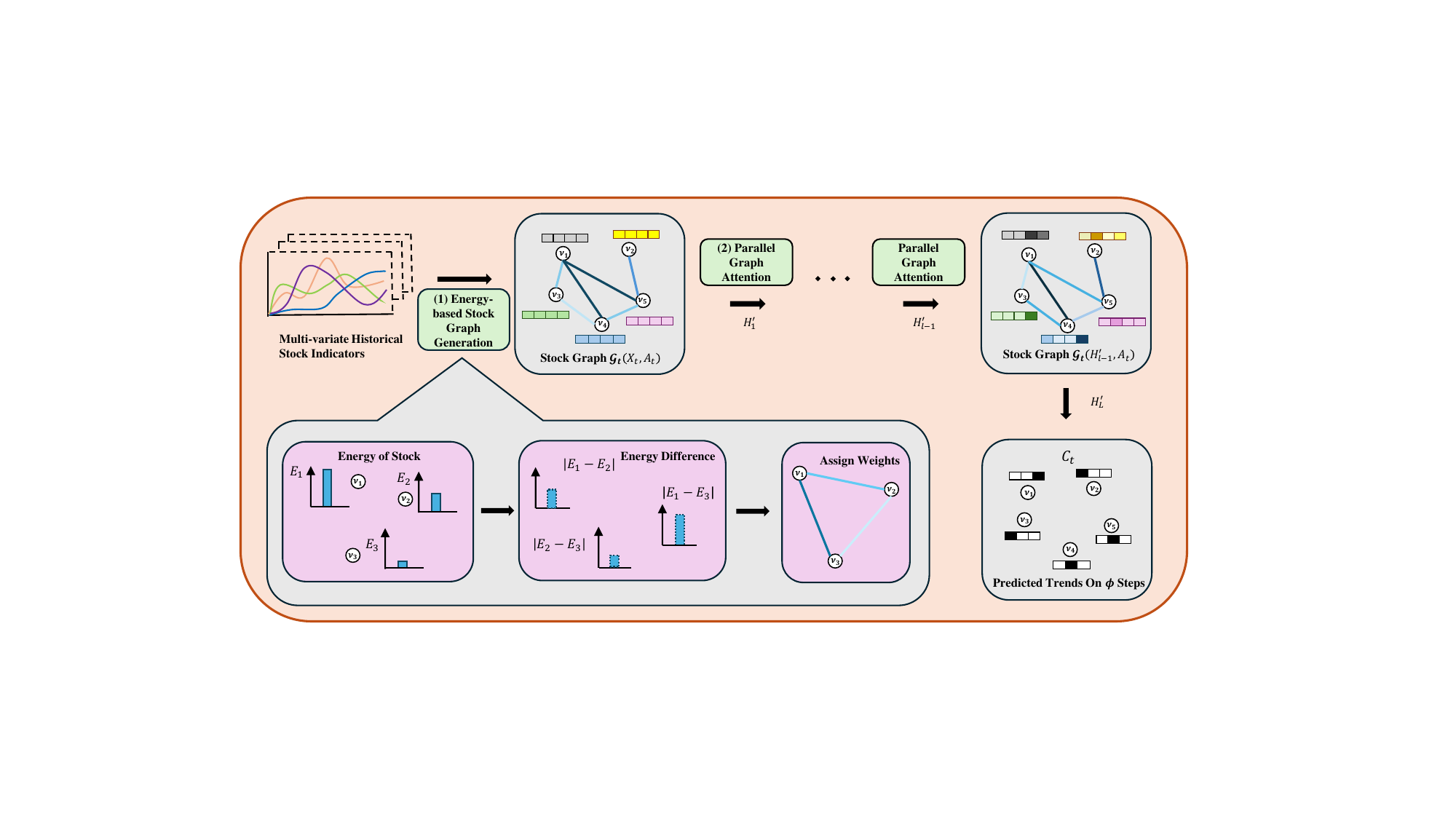}
\caption{The overall architecture of EP-GAT is structured as follows:  1) utilizing multiple historical stock indicators time series to construct the stock graph $\mathcal{G}_t(\mathbf{X}_t, \mathbf{A}_t)$, where the adjacency matrix $\mathbf{A}_t$ is generated by energy differences between stocks and the Boltzmann Distribution. It is designed to capture the evolving inter-dependencies between stocks; 2) applying the parallel graph attention blocks on the generated stock graph $\mathcal{G}_t(\mathbf{X}_t, \mathbf{A}_t)$, which learns the latent representation $\mathbf{H}'_l$ and preserves hierarchical temporal features within stocks; after $L$ blocks learning process the final representation $\mathbf{H}'_L$ is obtained. Then, a linear reshaping layer is applied to generate the predicted label matrix $\mathbf{C}_t$, which represents the stock movements on future $\phi$ steps.}
  \label{fig1}
\end{figure*}

\subsection{Boltzmann Distribution}
\indent
The Boltzmann distribution is a fundamental concept in statistical mechanics~\cite{landau2013statistical}. It describes the probability distribution of states in a particular system, where each state is weighted by a function of its energy. The distribution is defined as the following form,
\begin{equation}
    p_i = \frac{\mathrm{e}^{-\frac{E_i}{kT}}}{\sum_{o=1}^{M} \mathrm{e}^{-\frac{E_o}{kT}}}.
    \label{boltzman}
\end{equation}
Here, $p_i$ denotes the probability of a certain state, $E_i$ denotes the energy of the corresponding state, $M$ denotes the number of possible states, $k$ denotes the scaling factor, and $T$ denotes the absolute temperature of the system. It reflects the probability of a system occupying a certain state such that lower energy states are more likely to occur at lower temperatures and higher energy states become more accessible as temperature increases. 

\section{Methodology}
In this section, we elaborate on the detailed structure of the proposed EP-GAT, as depicted in Fig.\ref{fig1}.

\subsection{Energy-based Stock Graph Construction}
\indent
In terms of GNN-based approaches, the graph structure is crucial for effectively capturing interdependencies between nodes and learning meaningful representations~\cite{horn2021topological}. However, most existing GNN-based methods often adopt invariant relations to measure such complex interdependencies, such as HGNN~\cite{xu2022hgnn} constructing the graph according to the coexistence relationship, companies in the same industry are considered as connected, and LSTM-RGCN~\cite{li2021modeling} that adopt the realistic connection relationship, which means cross-shareholding companies are connected. These works contradict the stochastic nature of stock markets, where interactions between stocks are intricate and influenced by various time-varying aspects~\cite{guan2022dynagraph,qian2024mdgnn}. Hence, we propose an energy-based stock graph generation to address such a limitation. According to previous studies~\cite{shahzad2018global, ferrer2018time}, the energy of historical stock indicators time series can reflect their potential influences posed on other stocks. Moreover, prior works~\cite{sawhney2021exploring, arora2006financial} indicate that stocks with lower energy tend to be affected by stocks with higher energy. Meanwhile, \cite{wang2017network, kleinert2007boltzmann} demonstrate that when modelling multiple stocks as a system, their interactions can be approximated by the Boltzmann distribution. 

\indent
Accordingly, we can leverage energy differences between the historical stock indicator series and the Boltzmann distribution to capture the evolving inter-stock relations. The energy of a stock is defined by,
\begin{equation}
    E_i = \sum_{n=1}^{\tau F} ({\mathbf{X}_t^{i,n}})^2.
    \label{sig_energy}
\end{equation}
Here, $\mathbf{X}_t^{i,n}$ denotes the $n$-th element of the historical feature vector of the $i$-th stock at time step $t$.

Following \cite{de2011market}, the historical lag window size can be regarded as an approximation of the system temperature. Hence, we can utilize the lag window size $\tau$ as a direct approximation of absolute system temperature $T$, such that $T = \tau$. Moreover, we set $k$ as a hyperparameter to control the contribution of the energy difference. According to \eqref{boltzman} and \eqref{sig_energy}, the entry $\mathbf{A}^{i,j}_t$ is defined as,
\begin{equation}
    \mathbf{A}_t^{i,j} = \frac{\mathrm{e}^{-\frac{|E_i - E_j|}{k \tau}}}{\sum_{o=1}^{N} \mathrm{e}^{-\frac{|E_i - E_o|}{k \tau}}}.
    \label{adj_def}
\end{equation}
Here, $\mathbf{A}_t^{i,j}$ denotes the normalised similarity between stock $i$ and stock $j$ at time $t$, which is computed based on their energy difference. Accordingly, a large $\tau$ uniforms the influences of energy differences between stocks, and a small $\tau$ emphasises the influences of energy differences between stocks. This aligns with former findings~\cite{de2011market} that the potential influence of stocks tends to degrade as the period increases, while these effects are more distinguishable in shorter periods. In this sense, the complex interactions of stocks can be approximated by reweighing the probabilities at different states within the lag window. 

\begin{figure*}[tbhp]
    \centering
    \includegraphics[scale=0.65, trim= 4.65cm 4cm 4.5cm 4.2cm, clip]{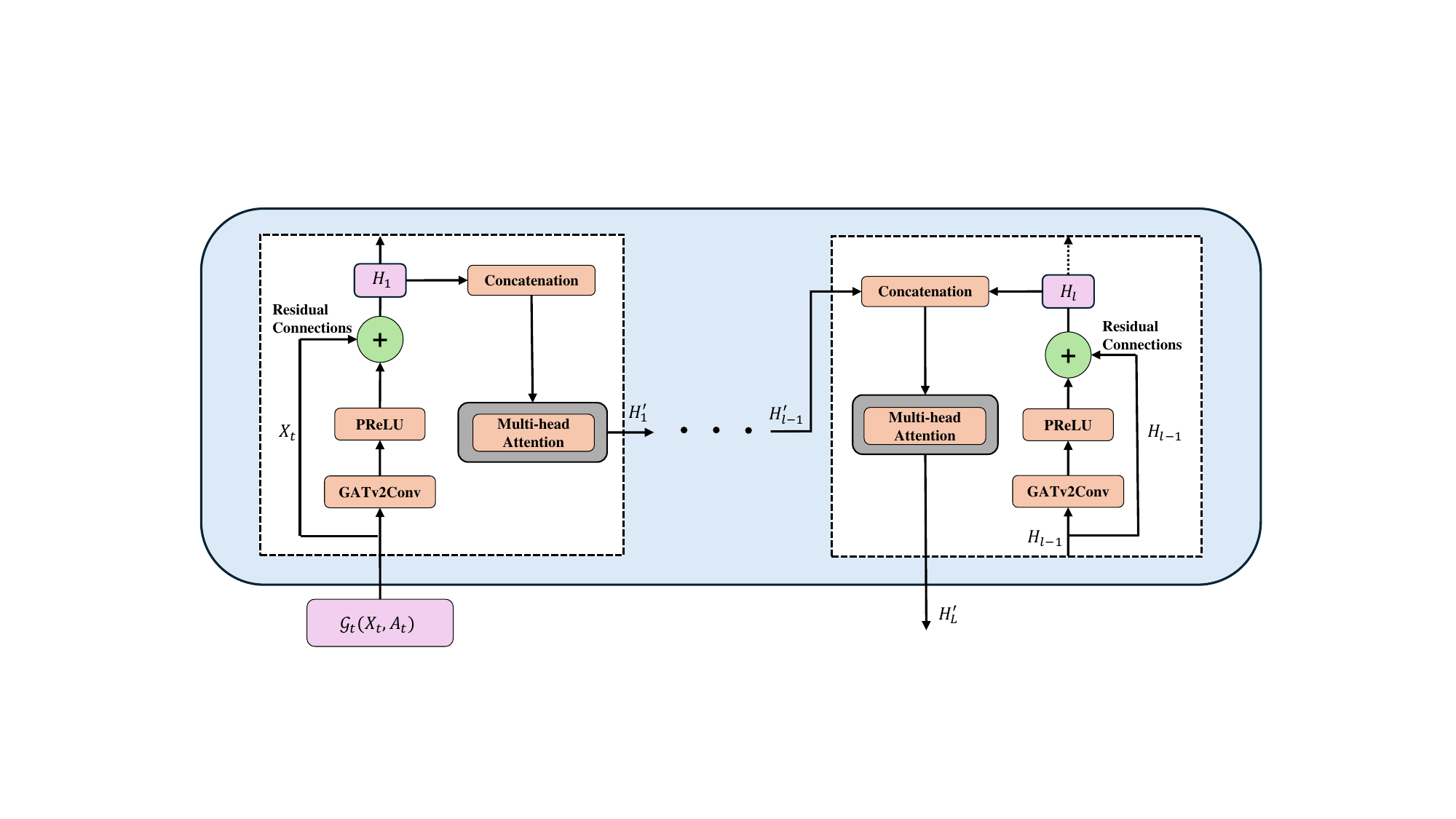}
    \caption{Detailed illustration of the parallel graph attention mechanism. Here, $\mathbf{H}_l$ denotes the latent representation from propagation, and $\mathbf{H}'_{l}$ denotes the latent representation matrix from the parallel graph attention. The concatenation is performed along the columns, and the GATv2Conv represents the layer-wise graph attention operation from~\cite{brody2021attentive}.}
    \label{fig2}
\end{figure*}

\indent
Besides, as~\cite{calandriello2018improved, chakeri2016spectral,gasteiger2019diffusion} indicates sparsifying graphs generated based on defined functions can improve the representation learning. This is often achieved by the threshold method and entry clipping method~\cite{gasteiger2019diffusion, sahu2021rethinking}. In this work, we adopt the threshold method by introducing a threshold value $s$ to sparsify the generated adjacency matrix $\mathbf{A}_t$, as defined below,
\begin{equation}
\mathbf{A}_t^{i,j} = \begin{cases} 
      \mathbf{A}_t^{i,j}, & \text{if } \mathbf{A}_t^{i,j} \geq s \\
      &\\ 
      0, & \text{else}.
   \end{cases}
   \label{sparsify}
\end{equation}

\subsection{Parallel Graph Attention}
\indent
According to~\cite{mantegna1999hierarchical}, stock markets exhibit hierarchical patterns, which reflect temporal changes of stocks on different scales~\cite{sawhney2021exploring}. However, most works often overlook this point. Furthermore, as information propagation continues, GNN-based methods can distort these hierarchical representations during the propagation~\cite{huang2020tackling,you2024multi}. Consequently, we proposed the parallel graph attention mechanism to capture the hierarchical features of stocks, which preserves and transforms the learned representation from each propagation layer, thereby preventing the distortion of hierarchical intra-stock features.

\indent
The parallel graph attention mechanism is defined as,
\begin{align}
\mathbf{H}'_{l} &=\gamma \Bigl(\mathbf{H}'_{l-1} ||(\eta(\mathbf{H}_{l-1})+ \mathbf{H}_{l-1}\mathbf{W}_l)\Bigr).
\label{eq8}
\end{align}
Here, $\mathbf{H}'_{l}$ denotes the learned node representation from the parallel graph attention mechanism, $\mathbf{H}_l$ denotes the intermediate node representation from propagation, $||$ denotes the concatenation operation over the column, $\eta(\cdot)$ denotes the GATv2Conv~\cite{brody2021attentive} propagation operation, and $\gamma(\cdot)$ denotes the multi-head attention. The multi-head attention is expressed as,
\begin{align}
    \mathbf{Q} &= \mathbf{H}_{l-1} \mathbf{W}_Q, \; 
    \mathbf{K} = \mathbf{H}_{l-1} \mathbf{W}_K, \; 
    \mathbf{V} = \mathbf{H}_{l-1} \mathbf{W}_V, \notag \\
    \text{head}_i &= \mathrm{softmax}\Bigl(\frac{\mathbf{Q}\,\mathbf{W}_{Q}^{(i)}\,(\mathbf{K}\,\mathbf{W}_{K}^{(i)})^\mathsf{T}}{\sqrt{d'}}\Bigr)\,
    \mathbf{V}\,\mathbf{W}_{V}^{(i)}\Bigr), \notag \\
    \mathbf{H}'_{l-1} &= \Bigl(\text{head}_1 \,\|\, \text{head}_2 \,\|\cdots\|\, \text{head}_h\Bigr) \,\mathbf{W}_O.
    \label{pgatdef}
\end{align}
Here, $h$ represents the number of attention heads, and $\mathbf{W}$ denotes the learnable matrices. From Fig.\ref{fig2}, we can observe that the representation learning and the information propagation are decoupled. This ensures the learned hierarchical features are not distorted during information propagation~\cite{zeng2021decoupling}, instead, they are preserved and re-transformed via multi-head attention.

\begin{table}[t]
\renewcommand\arraystretch{1.35}
    \centering
    \caption{Detailed statistics of LSE123, FTSE99, NYSE82, NASDAQ98 and SP101.}
    \resizebox{\linewidth}{!}{
    \begin{tabular}{c| c c c c c}
        \hline
        \hline
        \textbf{} & \textbf{LSE123} & \textbf{NYSE82} & \textbf{SP101} & \textbf{FTSE99} & \textbf{NASDAQ98} \\
        \hline
        \textbf{Train Period} & 07/2021-04/2023 & 07/2021-04/2023 & 07/2021-04/2023 & 07/2021-04/2023 & 07/2021-04/2023 \\
        \hline
        \textbf{Validation Period} & 04/2023-07/2023 & 04/2023-07/2023 & 04/2023-07/2023 & 04/2023-07/2023 & 04/2023-07/2023 \\
        \hline
        \textbf{Test Period} & 07/2023-07/2024 & 07/2023-07/2024 & 07/2023-07/2024 & 07/2023-07/2024 & 07/2023-07/2024 \\
        \hline
        \textbf{Train:Validation:Test} & 457 : 63 : 261 & 457 : 63 : 261 & 457 : 63 : 261 & 457 : 63 : 261 & 457 : 63 : 261 \\
        \hline
        \textbf{Number of Stocks ($N$)} & 123 & 82 & 101 & 99 & 98 \\
        \hline
        \textbf{Number of Indicators ($F$)} & 4 & 4 & 4 & 4 & 4 \\
        \hline
        \textbf{Number of Steps ($\phi$)} & 1 & 1 & 1 & 1 & 1 \\
        \hline
        \textbf{Number of Trends ($\alpha$)} & 2 & 2 & 2 & 2 & 2 \\
        \hline
        \hline
    \end{tabular}%
    }
    \label{table1}
\end{table}

\subsection{Objective Function}\label{E}
\indent
 According to \eqref{mapping} and following previous works~\cite{sawhney2021exploring,you2024dgdnn,you2024multi}, we perform the trend classification for future trading days. Thus, the objective function $L$ is formulated as follows,
\begin{equation}
    L = -\frac{1}{P} \sum_{t=1}^{T} \sum_{n=1}^{N} \sum_{i=1}^{\alpha} C_{t}^{n,i} \log(\hat{C}_{t}^{n,i}).
    \label{objective}
\end{equation}
Here, $P$ denotes the training period, $C_t^{n,i} = 1$ if stock $n$ at time $t$ belongs to class $i$, where $i=0$ denotes a downward trend and $i=1$ denotes an upward trend. $\hat{C}_{t}^{n,i}$ denotes the predicted label by the model. 

\section{Experiment}

\begin{table*}[th]
\renewcommand\arraystretch{1.8}
\caption{The performance comparison regarding ACC, MCC, and F1-Score of EP-GAT and five baseline models on the next trading day stock trend classification over test periods. The results are averaged on 10 runs with \textbf{bold} values indicating the best performance.}
\centering
\resizebox{\linewidth}{!}{
\begin{tabular}{c| c c c| c c c| c c c| c c c| c c c}
  \hline
  \hline
   \multirow{2}{*}{Method} & \multicolumn{3}{c|}{LSE123} & \multicolumn{3}{c|}{NYSE82} & \multicolumn{3}{c|}{SP101} & \multicolumn{3}{c|}{FTSE99} &\multicolumn{3}{c}{NASDAQ98}\\

   & ACC(\%) & MCC($\times10^{-2}$) & F1-Score  & ACC(\%) & MCC($\times10^{-2}$) & F1-Score & ACC(\%) & MCC($\times10^{-2}$) & F1-Score & ACC(\%) & MCC($\times10^{-2}$) & F1-Score & ACC(\%) & MCC($\times10^{-2}$) & F1-Score\\ 
\hline
    GraphWaveNet~\cite{wu2019graph} & $52.67\pm0.17$  & $-0.73\pm0.02$ & $0.53\pm0.02$ & $53.42\pm0.23$ & $-2.57\pm0.01$ & $0.54\pm0.01$ & $52.91\pm0.22$ & $-2.39\pm0.01$ & $0.52\pm0.01$ & $53.09\pm0.14$ & $1.77\pm0.02$ & $0.53\pm0.01$ & $53.17\pm0.19$ & $-3.44\pm0.02$ & $0.52\pm0.01$ \\
    HyperStockGAT~\cite{sawhney2021exploring} & $50.44\pm0.23$ & $1.03\pm0.01$ & $0.43\pm0.02$  & $50.39\pm0.31$ & $0.78\pm0.02$ & $0.46\pm0.01$ & $50.60\pm0.14$  & $1.30\pm0.01$ & $0.46\pm0.01$ & $50.24\pm0.16$ & $0.43\pm0.02$ & $0.45\pm0.03$ & $50.30\pm0.25$ & $0.58\pm0.02$ & $0.46\pm0.01$\\
    STGCN~\cite{yu2017spatio} & $51.22\pm0.24$ & $2.44\pm0.03$ & $0.51\pm0.01$ & $50.82\pm0.18$ & $1.64\pm0.03$ & $0.51\pm0.02$ & $51.14\pm0.20$ & $2.27\pm0.01$ & $0.51\pm0.02$ & $50.41\pm0.21$ & $0.81\pm0.02$ & $0.50\pm0.01$ & $50.62\pm0.24$ & $1.23\pm0.02$ & $0.51\pm0.02$\\
    Informer~\cite{zhou2021informer} & $53.23\pm0.14$ & $0.15\pm0.01$ & $0.52\pm0.01$ & $54.29\pm0.13$  & $2.44\pm0.02$ & $0.55\pm0.01$ & $53.81\pm0.16$ & $0.19\pm0.02$ & $0.52\pm0.01$ & $53.18\pm0.15$ & $0.29\pm0.02$ & $0.54\pm0.02$ & $53.12\pm0.18$ & $-0.40\pm0.01$ & $0.54\pm0.02$ \\
    ST-TIS~\cite{li2022lightweight} & $52.43\pm0.34$ & $0.51\pm0.03$ & $0.51\pm0.03$ & $52.31\pm0.17$ & $2.49\pm0.02$ & $0.53\pm0.01$ & $52.60\pm0.25$ & $-1.47\pm0.02$ & $0.50\pm0.03$ & $52.13\pm0.18$ & $0.14\pm0.02$ & $0.49\pm0.01$ & $52.46\pm0.36$ & $-0.06\pm0.02$ & $0.51\pm0.01$\\
  \hline
  EP-GAT & $\textbf{57.81}\pm\textbf{0.17}$  & $\textbf{2.45}\pm\textbf{0.02}$ & $\textbf{0.58}\pm\textbf{0.02}$  & $\textbf{56.68}\pm\textbf{0.27}$  & $\textbf{3.14}\pm\textbf{0.02}$ & $\textbf{0.57}\pm\textbf{0.01}$ & $\textbf{55.77}\pm\textbf{0.13}$  & $\textbf{2.32}\pm\textbf{0.02}$ & $\textbf{0.56}\pm\textbf{0.02}$  & $\textbf{55.09}\pm\textbf{0.17}$ & $\textbf{2.34}\pm\textbf{0.02}$ & $\textbf{0.57}\pm\textbf{0.03}$ & $\textbf{54.65}\pm\textbf{0.20}$ & $\textbf{4.80}\pm\textbf{0.03}$ & $\textbf{0.55}\pm\textbf{0.01}$ \\
  \hline
  \hline

  \end{tabular}%
}
\label{table2}
\end{table*}

\begin{table}[thp]
    \centering
    \caption{Ablation study configuration.}
    \begin{tabular}{l|cccc}
        \toprule
        Modules& EP-GAT & M1 & M2 & M3 \\
        \midrule
        Energy-based Stock Graph & \cmark & \cmark  & \xmark  & \xmark  \\
        Pre-defined Stock Graph & \xmark  & \xmark & \cmark & \cmark  \\
        GATv2Conv w/o Parallel Graph Attention & \xmark & \cmark  & \cmark & \xmark \\
        GATv2Conv w/ Parallel Graph Attention & \cmark & \xmark & \xmark & \cmark\\
        \bottomrule
    \end{tabular}
    \label{ablation_config}
\end{table}

\subsection{Experimental Setup}
\subsubsection{Dataset}
The datasets are, LSE123, FTSE99, NYSE82, NASDAQ98, and SP101, which are collected from index composites of the London Stock Exchange (LSE), Financial Times Stock Exchange (FTSE), New York Stock Exchange (NYSE), National Association of Securities Dealers Automated Quotations (NASDAQ), and Standard and Poor's 100 (S\&P 100). The numerical suffix represents the number of stocks. Stock indicators include open price, high price, low price, adjusted close price, and trading volume. The detailed dataset information is provided in Tab.\ref{table1}.

\subsubsection{Model Setting}
\indent
In this work, the lag window size $\tau$ is searched over $[7, 27]$, the constant $k$ is searched from $[0.02, 2]$, and the sparsification threshold $s$ is searched from $[0.25, 0.85]$. The number of stock indicators $F$ is set to 4. The number of heads $h$ is searched over $[2,30]$, and the number of layers $L$ is searched from $[2, 6]$. In addition, we adopt the PReLU as the activation function, and the AdamW as the optimizer. The learning rate is searched from $[10^{-4},2\times 10^{-3}]$ weight decay is searched from $[10^{-4},10^{-3}]$, and training epoch is set to 800. In this work, the number of trends $\alpha$ is set to 2, and forecast step $\phi$ is set to 1 following previous works \cite{sawhney2021exploring,cheng2021modeling,you2024dgdnn,you2024multi}, meaning a classification task of predicting next-day stock movements.

\subsubsection{Evaluation Metric}
Following previous studies \cite{wu2019graph,sawhney2021exploring,yu2017spatio,zhou2021informer,li2022lightweight}, we utilize three evaluation metrics for evaluation on the downstream task: classification accuracy (ACC), Matthews correlation coefficient (MCC), and F1-Score.

\subsection{Baseline Method}
\subsubsection{GNN-based Baseline}
\begin{itemize}
 \item STGCN~\cite{yu2017spatio}: A spatial-temporal convolutional network operates on graphs, which focuses on the nonlinearity and complex inter-relations in traffic flow forecast. The interdependencies between different areas are constructed based on the distance between stations and keep updating over time.
 \item GraphWaveNet~\cite{wu2019graph}: A spatial-temporal graph modelling approach that captures latent spatial-temporal interdependencies between multiple time series by integrating an adaptive dependency matrix and stacked dilated convolutions.
    \item HyperStockGAT~\cite{sawhney2021exploring}: A graph learning method that models complex intra-stock relations by adopting a hyperbolic graph attention on Riemannian manifolds. It applies a static stock graph generation method, companies within the same sector are connected.
\end{itemize}
\subsubsection{Transformer-based Baseline}
\begin{itemize}
    \item Informer~\cite{zhou2021informer}: A Transformer-based model for multivariate time series modelling. Concretely, it learns long-range dependencies of multiple time series with the ProbSparse self-attention, and self-attention distilling to handle extreme series. Furthermore, it adopts a generative decoder for rapid inference across extended periods.
    \item ST-TIS~\cite{li2022lightweight}: A Transformer-based method fuses latent representations from sampled regions in traffic flow predictions, enabling dynamic modelling of complex spatial-temporal dependencies between areas (nodes).
\end{itemize}

\subsection{Performance Evaluation}

\begin{figure*}[tbp]
    \centering
    \includegraphics[width=\linewidth]{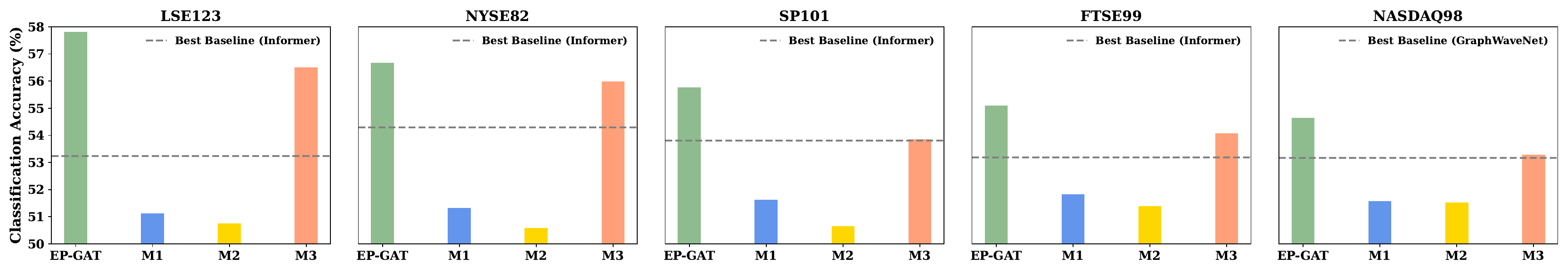} 
    \caption{The EP-GAT comprises two components: energy-based stock graph generation and parallel graph attention mechanism. The ablation studies are conducted by replacing energy-based stock graph generation with the pre-defined stock graph or removing the parallel graph attention mechanism on the backbone GATv2Conv. The grey dot lines denote the best-performing baseline models.}
    \label{fig3}
\end{figure*}
\indent 
The performance evaluation results on the next-day stock trends classification are shown in Tab.\ref{table2}. From these results, we can observe that the proposed model EP-GAT consistently outperforms five baselines with a large margin over test periods across all metrics. Specifically, EP-GAT achieves an average improvement of 7.61\% in ACC, $2.63 \times 10^{-2}$ in MCC, and $0.06$ in F1-Score compared to baseline models. Furthermore, the following observation can be made based on the experimental results.

On the one hand, among the GNN-based models, methods (GraphWaveNet and EP-GAT) that capture dynamic interdependencies outperform other GNN-based models (HyperStockGAT and STGCN). This highlights the significance of dynamic modelling in inter-stock relations. On the other hand, other GNN-based models (HyperStockGAT, STGCN, GraphWaveNet) that do not consider the hierarchical intra-stock features perform worse than EP-GAT. This validates the effectiveness of the parallel graph attention mechanism in learning the hierarchical intra-stock features. Furthermore, Transformer-based baselines (Informer and ST-TIS) outperform GNN-based baselines using static stock graphs (HyperStockGAT, and STGCN). Specifically, Transformer-based methods realize 52.96\% in ACC, $0.43 \times 10^{-2}$ in MCC, and 0.52 in F1-score on average over five datasets. While GNN-based baselines that apply static stock graphs achieve 50.62\% in ACC, $1.25\times 10^{-2}$ in MCC, 0.48 in F1-score on average over five datasets. These observations verify the significance of capturing the evolving inter-stock relations and preserving the hierarchical intra-stock features.

\subsection{Ablation Study}
\indent
To verify the effectiveness of the proposed components in EP-GAT, we conduct ablation studies by substituting energy-based stock graph generation with the pre-defined industry stock graph~\cite{sawhney2021exploring} or removing parallel graph attention from backbone GATv2. The configurations of ablation studies are shown in Tab.\ref{ablation_config}, and ablation results are shown in Fig.\ref{fig3}. We compare the performance on stock trend classification between EP-GAT and the other three variants (M1, M2, and M3) composed of different modules. Specifically, compared to EP-GAT, the performance of M3 degrades by 2.25\% on average. This highlights the importance of considering dynamic inter-stock relations in the modelling of the stock markets. Furthermore, it reiterates the effectiveness of adopting well-designed kernel functions to capture the complex interdependencies between stocks, aligning with their stochastic and evolving nature~\cite{kasa1992common, adam2016stock}. Then, upon removing the parallel graph attention mechanism, there is a reduction in performance by 8.05\%, as shown in comparisons between EP-GAT and M1. Meanwhile, a decrease in performance by 6.86\% is presented in the comparison between M2 and M3. These illustrate the parallel graph attention's effectiveness in capturing long-term temporal representations, which facilitates the modelling of the hierarchical intra-stock features.

\begin{figure}[htbp]
    \centering
    \includegraphics[width=0.4\textwidth]{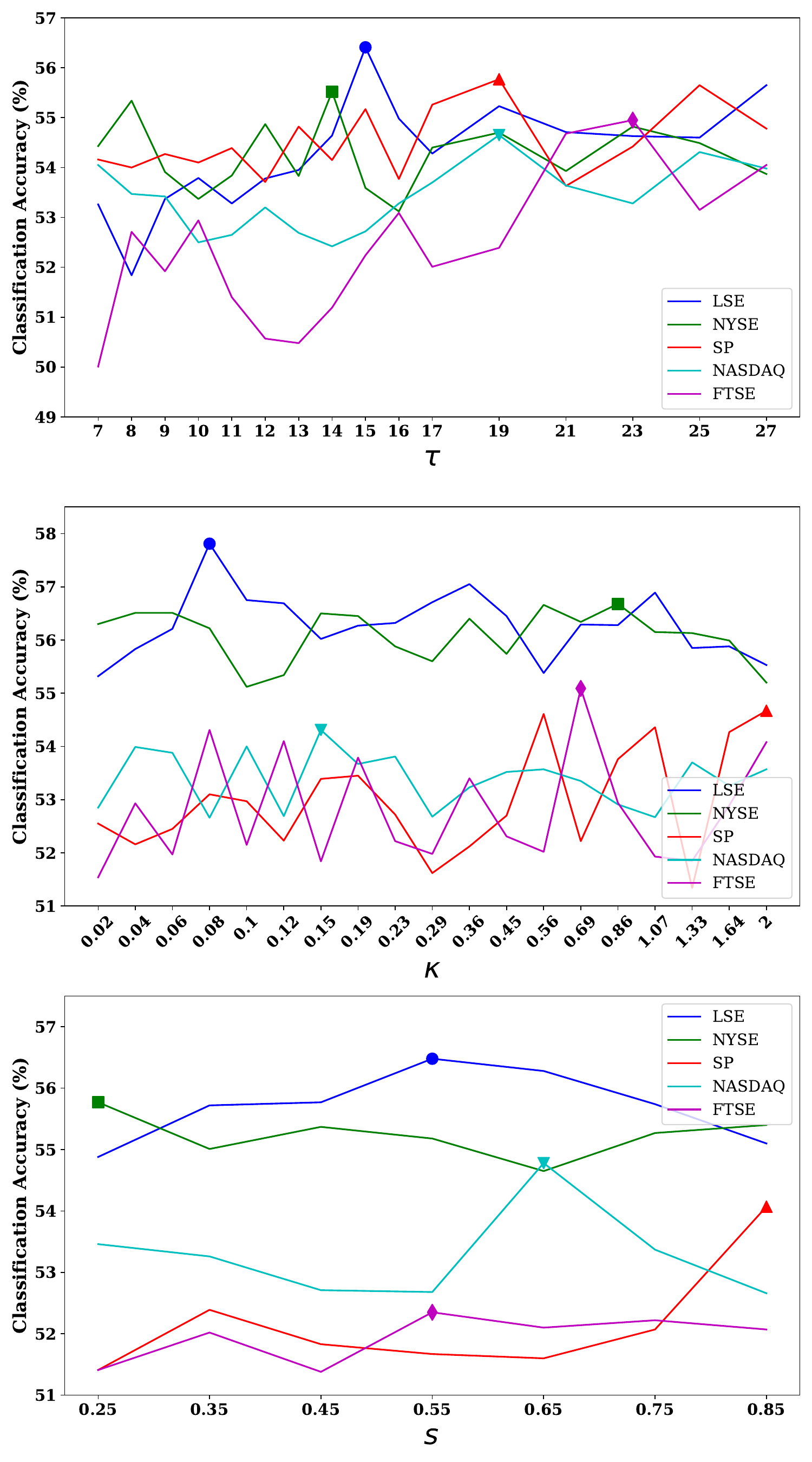}
    \caption{Hyperparameter sensitivity results of the lag window size $\tau$, scaling factor $k$ and threshold $s$ on five datasets over test periods.}
    \label{fig4}
\end{figure}

\subsection{Hyperparameter Study}
\indent 
In this section, we conduct the sensitivity analysis of three critical hyperparameters in this work: the lag window size $\tau$, the scaling factor $k$, and the sparsification threshold $s$. Experimental results are shown in Fig.\ref{fig4}.

\subsubsection{Lag Window Size $\tau$}
From Fig.\ref{fig4}, we can observe that EP-GAT reaches the optimal performance over five datasets when $\tau \in [14, 23]$. This coincides with prior findings~\cite{adam2016stock,you2024dgdnn} that a lookback window of around 20 days yields better results in stock trend forecasting tasks. Moreover, if $\tau$ does not belong to this interval, then the model performance generally degrades and fluctuates. This indicates that the lag window should be neither too large nor too small, such that the model can effectively learn from historical stock indicators, which should contain sufficient temporal features characterizing stocks.

\subsubsection{Scaling Factor $k$}
\indent
As defined in \eqref{boltzman}, when the scaling factor $k$ increases, it gradually eliminates differences between assigned weights in the adjacency matrix $\mathbf{A}_t$. Based on the results in Fig.\ref{fig4}, the optimal value of $k$ varies across different datasets. Concretely, EP-GAT  achieves the optimal performance on LSE123 when $k = 0.08$ and on NASDAQ98 when $k = 0.15$. While for the other three datasets, the optimal performance is reached when $k \in \{0.56, 0.69, 0.86\}$. These imply that stocks in the LSE123 and NASDAQ98 have more intricate and distinctive interdependencies~\cite{jiang2011comparison, schwert2002stock} than the other three, reflected as significant differences in the weights of entries.  

\subsubsection{Sparsification Threshold $s$}
\indent
According to \eqref{sparsify}, the threshold $s$ is utilized to sparsify the adjacency matrix. When $s \in \{0.55, 0.65, 0.85\}$, the model consistently demonstrates strong performance. This suggests that threshold $s$ should be neither too large nor too small.  In general, smaller threshold values result in dense adjacency matrices, which can introduce noisy signals and interference. Therefore, dense adjacency matrices can pose negative effects on the model performance. In contrast, larger threshold values lead to sparse matrices that remove the majority of connections between stocks. This limits the representation learning from the stocks, as most interdependencies between stocks are eliminated with large threshold values. Accordingly, carefully balancing the sparsification threshold $s$ is critical for the model to effectively capture complex inter-stock relations and learn from the generated stock graphs.

\section{CONCLUSION}
\indent 
In this work, we present EP-GAT, a novel framework for predicting future trends of multiple stocks. To address the limitation of existing works adopting static factors to model evolving inter-stock relations, we propose energy-based stock graph generation. This is achieved by considering energy differences between stocks within a lag window, which are then re-weighted based on the Boltzmann distribution. Then, we present a parallel graph attention mechanism to preserve hierarchical intra-stock features by gradually fusing latent representations from each propagation layer via multi-head attention. Experimental results show that EP-GAT consistently outperforms five baseline models over five datasets across three metrics. The results of ablation studies further validate the effectiveness of the proposed approach EP-GAT. 

Nonetheless, this work has two limitations. First, the generated stock graphs are undirected, which does not consider directed influences between stocks. In some real-world scenarios, less influential stocks tend to be affected by more influential stocks. Conversely, more influential stocks tend not to be affected by less influential stocks. Second, the parallel graph attention mechanism involves multi-head attention and feature-dependent operation (i.e., concatenation), which can lead to difficulties in training if the generated stocks are large. This might hinder EP-GAT from modelling the future trends of substantial stocks in a market. Future work will consider directed inter-stock relations and improving the parallel graph attention mechanism with the retention mechanism.

\bibliographystyle{ieeetr}
\bibliography{example}

\end{document}